\documentclass[aps,prx,twocolumn,superscriptaddress,showpacs]{revtex4-1}
\usepackage{amsmath,amssymb,wasysym,graphicx}

\usepackage[utf8]{inputenc}
\usepackage[T1]{fontenc}
\usepackage{xcolor}

\IfFileExists{newtxtext.sty}
   {\usepackage{newtxtext,newtxmath}}
   {\IfFileExists{stix.sty}
      {\usepackage{stix}}
      {\IfFileExists{mathptmx.sty}
      {\usepackage{mathptmx}}{} } }

\usepackage{textcomp}

\usepackage{bm}

\IfFileExists{siunitx.sty}{\usepackage{booktabs,siunitx}}{}

\pdfoutput=1
\usepackage{color}
\definecolor{LinkColor}{rgb}{0.256,0.439,0.588}
\usepackage{hyperref}
\hypersetup{
   pdfauthor={good guys},
   pdftitle={good title},
   colorlinks=true,
   citecolor=LinkColor,
   linkcolor=LinkColor,
   urlcolor=LinkColor
}

\renewcommand{\vec}[1]{\mathbf{#1}}

\usepackage{pifont}
\usepackage[normalem]{ulem}

\usepackage{multirow}

\begin{document}

\title{Kekul\'e valence bond order in an extended Hubbard model on the honeycomb lattice, with possible applications to twisted bilayer graphene}

\author{Xiao Yan Xu}
\affiliation{Department of Physics, Hong Kong University of Science and Technology, Clear Water Bay, Hong Kong, China}
\author{K. T. Law}
\affiliation{Department of Physics, Hong Kong University of Science and Technology, Clear Water Bay, Hong Kong, China}
\author{Patrick A. Lee}
\email{palee@mit.edu}
\affiliation{Department of Physics, Massachusetts Institute of Technology, Cambridge MA 02139, USA}

\date{Apr 27, 2018}

\begin{abstract}
Using large-scale quantum Monte Carlo simulations, we exactly solve a model of Fermions hopping on the honeycomb lattice with cluster charge interactions, which has been proposed as an effective model with possible application to twisted bilayer graphene near half-filling.
We find an interaction driven semimetal to  insulator transition to  an insulating phase consisting of a valence bond solid with Kekul\'e pattern.  Finite size scaling reveals that the phase transition of the semimetal to Kekul\'e valence bond solid phase is continuous and belongs to chiral XY universality class. 
\end{abstract}

\maketitle
Correlation driven metal to insulator transition provides a mechanism to generate insulators beyond the band  picture. 
One well known example is the undoped cuprates. According to band theory, it has a half-filled band and should be a metal, but the strong interaction of localized $d$ orbitals makes it a Mott insulator with antiferromagnetic long range order~\cite{lee2006doping}. More interestingly, the doped cuprates shows unconventional superconductivity which cannot be explained by traditional BCS theory~\cite{bardeen1957theory}.
The recently discovered twisted bilayer graphene at small "magic" angle is a new system of correlated insulator~\cite{cao2018correlated}. The twisted bilayer graphene forms moir\'e pattern, and  flat band emerges at some "magic" twist  angles, according to theoretical calculations~\cite{bistritzer2011moire,morell2010flat,santos2012continuum,fang2016electronic,trambly2012numerical}. The recent experiments on twisted bilayer graphene (TBG) near one of small "magic" angles found correlated insulator behavior and unconventional superconductivity by gating~\cite{cao2018correlated,cao2018unconventional}. 

To understand those exotic phases in TBG, the first task is to know the nature of the insulating phase. In experiments~\cite{cao2018correlated}, the conductance data show that near charge neutrality  the bands may contain nodes as the conductance shows a "V"-shaped dip. This is consistent with the moir\'e band picture where there are Dirac points at the  Brillouin zone (BZ) corner, separating two sets of band, each of which can accommodate 4 electrons including spin ~\cite{bistritzer2011moire,morell2010flat,santos2012continuum,fang2016electronic,trambly2012numerical}. The correlated insulating phase occurs at half filling of the electron or hole band, where the occupation number per superlattice unit cell is $\pm 2$.
 In addition, SdH oscillation measurements find Fermi pockets near half-filling with an area given by the doped electron or hole density,  suggesting the existence of a correlation induced gaps.
With all these observations in mind, what is the proper effective theory? As the charge center of the TBG forms a triangular lattice, it is tempting to start from orbitals on triangular lattice~\cite{xu2018topological,guo2018pairing,dodaro2018phases}. However, this cannot produce the Dirac nodes at the charge neutrality point mentioned above. Recently it has been proposed that  it is necessary to construct a model on a honeycomb lattice with 2 orbitals per site to account for the symmetry of the LDA band~\cite{po2018origin,yuan2018model}. While there are subtle differences on possible obstructions to the construction of localized Wannier orbitals, these papers are in agreement that the honeycomb lattice is the proper starting point for a tight-binding formulation. By making further assumptions about the breaking of valley symmetry, Po \emph{et~al.}~\cite{po2018origin} proposed the following Hamiltonian $H=H_t+H_U$ to describe the subset of hole (or electron) bands.  
\begin{equation}
H_t=-\sum_{ij}\sum_{\alpha} t_{ij} c_{i\alpha}^\dagger c_{j\alpha} + h.c.
\end{equation}
\begin{equation}
H_{U}=U\sum_{\hexagon} (Q_{\hexagon}-2)^2
\end{equation}
The model consists of a single orbital with spin degeneracy on the honeycomb lattice. Motivated by the fact that the charge is concentrated on the triangular lattice formed by the hexagonal plaquette, the interaction term punishes occupation of the cluster charge $Q_{\hexagon} \equiv \sum_{i \in \hexagon} \frac{n_i} {3}$ ( $n_i=\sum_{\alpha}  c_{i\alpha}^\dagger c_{i\alpha}$) when it deviates from 2 per plaquette. Note in this definition, the interaction strength is $\frac{2}{3} U$ for onsite, $\frac{4}{9}U$ for nearest-neighbor (NN), and $\frac 2 9 U$ for next-NN (NNN) and third-NN (the interaction strength ratio from onsite to third-NN is 3:2:1:1). Besides the difference in the definition of the onsite $U$, it is possible that the additional NN, NNN and third-NN repulsions add a mean field background and effectively reduce the strength of the onsite repulsion. This may explain why the semi-metal phase will turn out to be stable up to a larger value of $U/t$ compared with the standard Hubbard model. We shall study this model using QMC for half-filling, ie 2 electrons per unit cell, or a single electron per site. When $U$ is small we expect a semi-metal with Dirac spectrum and we shall look for the onset of an energy gap with increasing $U$. The cluster charge interaction term distinguishes this model from the conventional SU(2) Hubbard model on a honeycomb lattice which has been intensively studied, and known to exhibit a continuous transition from a semimetal (SM) to an AB sublattice anti-ferromagnetic insulator (AFMI) ~\cite{sorella2012absence,assaad2013pinning,otsuka2016universal}. As we shall see, the cluster charge model behaves very differently and is therefore of intrinsic interest, even if its applicability to twisted bilayer graphene remains to be established.

\begin{figure}[t!]
\includegraphics[width=0.9\columnwidth]{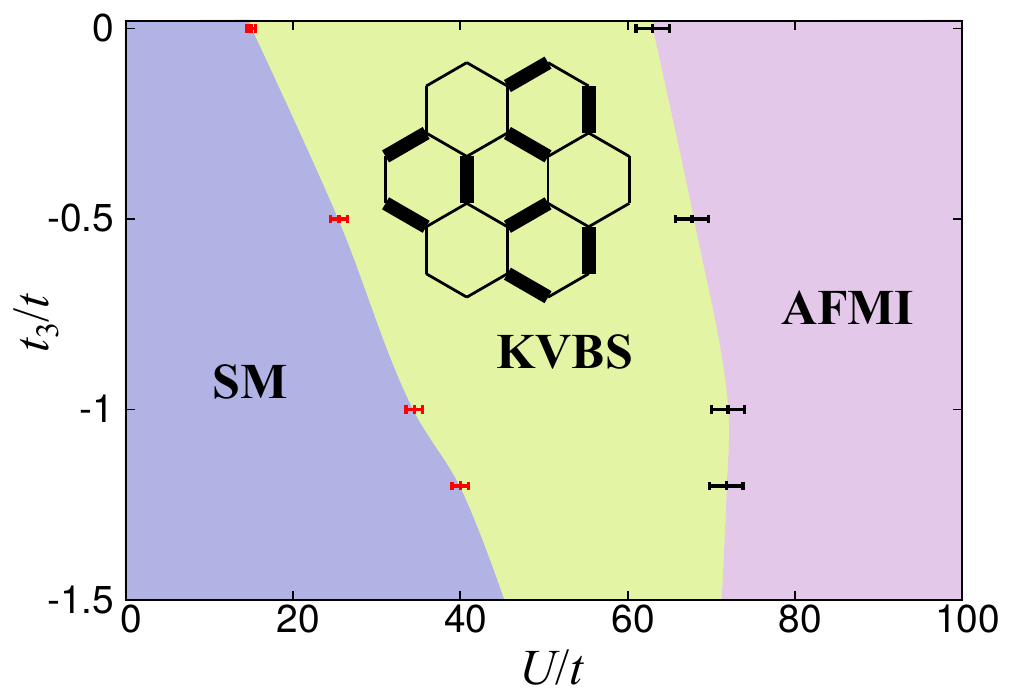}
\caption{$U/t-t_3/t$ phase diagram of a spinful model with first neighbor hopping $t$, third neighbor hopping $t_3$ and cluster charge interaction $U$ on honeycomb lattice.  SM denotes semimetal phase, KVBS denotes Kekul\'e valence bond solid phase, AFMI denotes AB sublattice antiferromagnetic insulating phase. The transition from SM to KVBS (red data points) is continuous and belongs to chiral XY universality class. The transition from KVBS to AFMI (black data points)  is first order.
}
\label{fig:phase-diagram}
\end{figure}

{\it Model and Method}\,---\,
We study the cluster charge model on the honeycomb lattice using QMC. At half-filling (2 electrons per unit cell), the simulation is sign-free as long as hopping is limited to between opposite sub-lattices. 
In the simulation, we will take first neighbor hopping $t$ as energy unit, and explore the phase diagram by varying cluster charge interaction strength $U/t$ and  third neighbor hopping $t_3/t$. For the method,  we use determinantal QMC~\cite{blankenbecler1981monte,hirsch1985two,assaad2008world}, which is a standard method to solve the interacting lattice model  when there is no sign problem. Particularly, to study the ground state properties, the projection version of determinantal QMC (PQMC)  is chosen.  PQMC starts with a trial wave function $|\Psi_T\rangle$, and the ground state wave function is obtained from projection by the time evolution operator  $|\Psi_0\rangle = e^{-\frac{\Theta}{2} H}|\Psi_T\rangle$ as $\Theta$ goes to infinity. Physical observables can be calculated as $\langle\hat{O}\rangle=\frac{\langle\Psi_{0}|\hat{O}|\Psi_{0}\rangle}{\langle\Psi_{0}|\Psi_{0}\rangle}$, or more explicitly, 
\begin{equation}
\langle\hat{O}\rangle=\frac{\langle\Psi_{T}|e^{-\frac{\Theta}{2}H}\hat{O}e^{-\frac{\Theta}{2}H}|\Psi_{T}\rangle}{\langle\Psi_{T}|e^{-\Theta H}|\Psi_{T}\rangle}
\end{equation}
 The projection time is divided into $M$ slices ($\Theta=M\Delta_\tau$). To treat  the interaction part $H_I$, we first perform a Trotter decomposition to separate $H_t$ and $H_U$ in exponential $e^{-\Delta_\tau H} = e^{-\Delta_\tau H_t}e^{-\Delta_\tau H_U} + \mathcal{O}(\Delta_\tau^2)$. A further Hubbard-Stratonovich (HS) transformation
 is performed on the interaction  part to get fermion bilinears coupled to discrete auxiliary fields. After tracing out the free fermions degrees of freedom, we can perform Monte Carlo sampling on the discrete auxiliary fields and measure the physical observables. We choose  
 the ground state wavefunction of half-filled non-interacting systems (described by $H_t$)
 as the trial wave function $|\Psi_T\rangle$. In the simulation, we set $\Theta=2L$ if not specified. For imaginary time slices, we set $\Delta_\tau=0.1$ for small $U$ and $t_3$, and $\Delta_\tau=0.05$ for large $U$ or $t_3$  ($U/t>30$ or  $|t_3/t|>0.5$).  We have tested that this setup is enough to get converged and error controllable results. For the sign problem free of the model we simulate, a concise argument  is that  as the model has particle-hole symmetry, the particle-hole transformation on  spin down fermion effectively makes the positive $U$ to be negative one, then we decouple the interaction part to density channel by HS transformation. Finally the Monte Carlo weight can be written as a square of determinant (spin up and down fermion have same determinant) of  matrix only with real numbers, which is always semipositive. More details of the PQMC formulation and the absence of sign problem can be found in Supplemental Material~\cite{suppl}. 

{\it Results}\,---\, 
The $U/t - t_3/t$ ground state phase diagram at half-filling is showed in Fig.~\ref{fig:phase-diagram}. We found three phases in total\,---\,they are semimetal (SM) phase which is connected to non-interacting case, the AFMI phase which is connected to the large $U$ limit, and a Kekul\'e valence bond solid (KVBS) phase which is new. Thus in contrast to the local Hubbard model, an intermediate KVBS phase appears in a large part of the phase diagram. Furthermore, surprisingly, the transition between SM and KVBS is continuous. 
On the other hand, the phase transition from KVBS to AFMI phase appears to be first order.

\begin{figure}[t!]
\includegraphics[width=0.9\columnwidth]{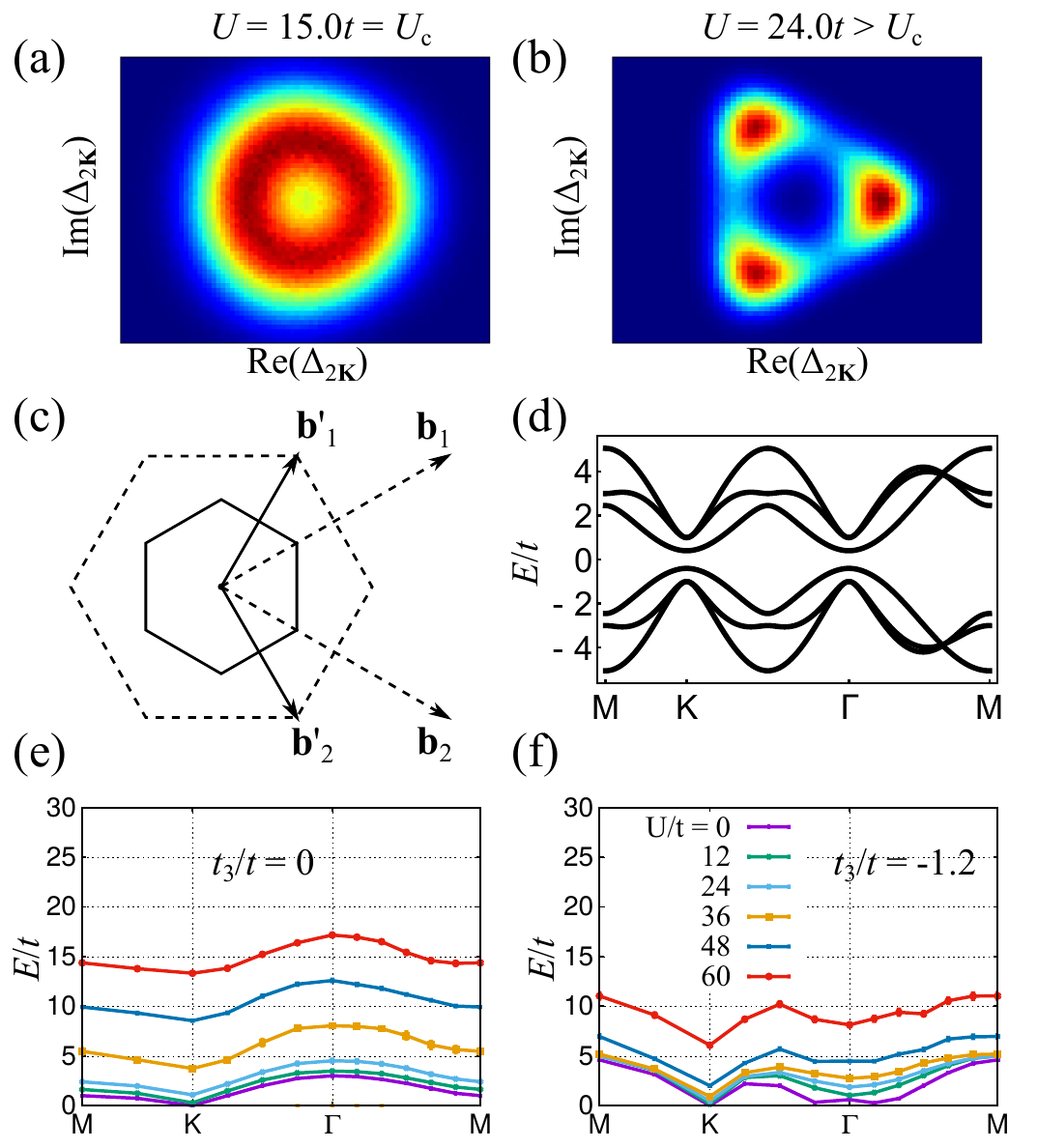}
\caption{(a) Histogram of KVBS order parameter at the SM to KVBS phase transition point and (b)  in the KVBS phase with $L=12$ and $t_3/t=0$. (c) Dashed hexagon denotes first BZ of the honeycomb lattice, so called large BZ in main text. Solid line hexagon denotes the folded BZ by KVBS order, called the small BZ in the main text.  (d) KVBS mean field band structure in $t_3/t=-1.2$ case with order  parameter $\Delta_{2\vec{K}}=2$. (e) Single particle gap for different momentum points obtained from QMC at $t_3/t=0$ and (f) at $t_3/t=-1.2$. Here the single particle gap is obtained from the exponential decay of the time-displaced Green's function of the  $L=12$ systems and more details are showed in Supplemental Material~\cite{suppl}.
}
\label{fig:kekuleorder}
\end{figure}

The KVBS order is shown in the inset of Fig.~\ref{fig:phase-diagram}. It breaks the lattice translational symmetry, resulting in  a $\sqrt{3}\times \sqrt{3}$ enlarging of unit cell. The broken symmetry is Z3, corresponding to the three ways to triple the unit cell size. Let us define $\vec{K}=\frac 1 3 \vec{b}_1 + \frac 1 3 \vec{b}_2$ where $\vec{b}_1$ and $\vec{b}_2$ are reciprocal lattice vectors as showed in Fig.~\ref{fig:kekuleorder}(c). $\vec{K}$ is the zone corner of the original (large) BZ. The KVBS order is formed by coupling electrons at $\vec{K}$ and $-\vec{K}$, so the order parameter can be defined as $\Delta_{2\vec{K}} = \sum_{i,\alpha}e^{i2\vec{K}\cdot \vec{r}_i}\left( c_{i,\alpha}^{\dagger} c_{i+\delta,\alpha} + \text{h.c.} \right)$.
The phase of $\Delta_{2\vec{K}}$ captures the $Z_3$ symmetry breaking, which is related by $2\pi/3$ rotation of the phase. This is demonstrated in Fig.~\ref{fig:kekuleorder}(b) which shows the histogram of the real and imaginary part of $\Delta_{2\vec{K}}$.
 In the KVBS phase, the histogram 
 clusters into three region   and they are connected by $C_3$ rotation as showed in  Fig.~\ref{fig:kekuleorder}(b). In the thermodynamic limit, one of the three branches  will be chosen and $C_3$ ($Z_3$) symmetry is broken.  

In literature, the KVBS phase was studied in several SU(N) quantum Heisenberg (-like) models and Hubbard models on honeycomb lattice~\cite{read1990spin,lang2013dimerized,li2017fermion,zhou2016mott,sato2017dirac}. The KVBS order was even taken as a mechanism to  realize charge fractionalization with time reversal symmetry in graphene~\cite{hou2007electron}.
The continuous transition between SM and KVBS is intuitively unexpected  because  a third order invariant of the KVBS order parameter can be constructed and according to Landau mean field theory the transition is predicted to be first order.  However, if the quantum fluctuations of gapless fermionic modes are considered,
the continuous transition is made possible~\cite{li2017fermion,zhou2016mott,scherer2016gauge,jian2017fermion,classen2017fluctuation,torres2018fermion}.
More interestingly, at the SM to KVBS transition point, both large-N renormalization group (RG) and functional RG calculations~\cite{li2017fermion,classen2017fluctuation,jian2017fermion} show that the $C_3$ rotational symmetry of the order parameter is enlarged to be a continuous one and the transition belongs to the chiral XY universality class, as opposed to the chiral-Heisenberg class for the SM to AFMI transition ~\cite{assaad2013pinning,otsuka2016universal,mihaila2017gross,zerf2017four}. The emergence of the U(1) symmetry is also found in our SM to KVBS transition, and are showed  in  Fig.~\ref{fig:kekuleorder}(a). 

\begin{figure}[t!]
\includegraphics[width=0.9\columnwidth]{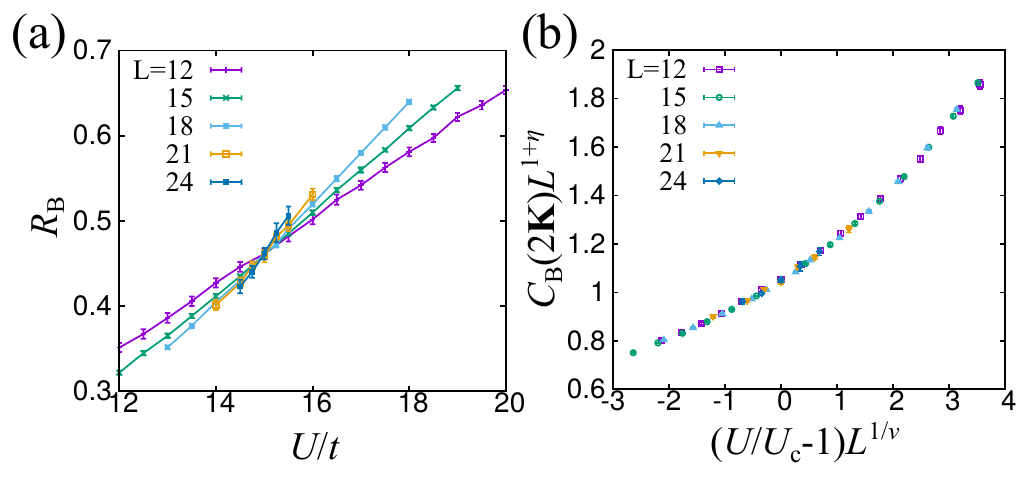}
\caption{(a) Correlation ratio of the bond correlation for $t_3/t=0$. The crossing point gives an estimate of the critical point $U_c=15.0(2)$.  (b) Data collapse of bond structure factor at momentum $2\vec{K}$, which is the absolute value squared of the KVBS order parameter. The transition from SM to KVBS belongs to chiral XY universality class. The data collapse gives $\nu=1.05(5)$, $\eta=0.76(2)$.
}
\label{fig:data-collapse}
\end{figure}

In order to  characterize the SM to KVBS transition, we measured the correlation ratio of the KVBS order $R_\text{B}(U,L) =1-\frac{C_\text{B}(2\vec{K}+\delta \vec{q})}{C_\text{B}(2\vec{K})}$ for different interaction $U$ and system size $L$, where $C_\text{B}(\vec{q})= \frac{1} {L^4} \sum_{i,j} e^{i\vec{q}\cdot (\vec{r}_i-\vec{r}_j)} \langle B_i B_j \rangle $  is structure factor of bond correlation with $\delta$-direction bond $B_i$ defined as $B_i=\sum_{\alpha} ( c_{i,\alpha}^{\dagger} c_{i+\delta,\alpha} + \text{h.c.} ) $.  In the above formula, $\delta \vec{q}$ is the smallest momentum of the lattice,  $\langle \cdots \rangle$ denotes Monte Carlo average and $2\vec{K}$ is the $Q$-vector of KVBS order, which connects different valleys at $\vec{K}$ and $-\vec{K}$. 
  The correlation ratio $R_\text{B}(U,L)$ is a renormalization invariant  quantity of the continuous SM to KVBS transition and it will cross at a point $U_c$ for different system size $L$. The crossing point $U_c$ gives an estimation of the location of the quantum critical point (QCP). 
As there is an emergent continuous U(1) at the QCP (see  Fig.~\ref{fig:kekuleorder}(a)), the critical behavior of the SM with chiral Dirac fermions to KVBS phase transition might be described by the chiral XY universality class~\cite{li2017fermion,zhou2016mott,scherer2016gauge,jian2017fermion,classen2017fluctuation,torres2018fermion,zerf2017four}.
To confirm this conjecture, 
we perform a finite size scaling of the KVBS structure factor near the QCP, and make a data collapse to find the critical exponents.
We assume Lorentz symmetry ($z=1$) here and expect $C_\text{B}(2\vec{K},U,L)L^{z+\eta} = f_\text{B} ( (U/U_c-1)L^{1/\nu} )$. The data collapse process is to find $\nu$ and $\eta$ to make all data points collapse at one single unknown curve described by function $f_\text{B}$, as showed in Fig.~\ref{fig:data-collapse}. 
 We find $\nu=1.05(5)$ and $\eta=0.76(2)$, which are comparable with the QMC results on different models~\cite{li2017fermion,otsuka2018quantum}.
 
\begin{figure}[t!]
\includegraphics[width=0.9\columnwidth]{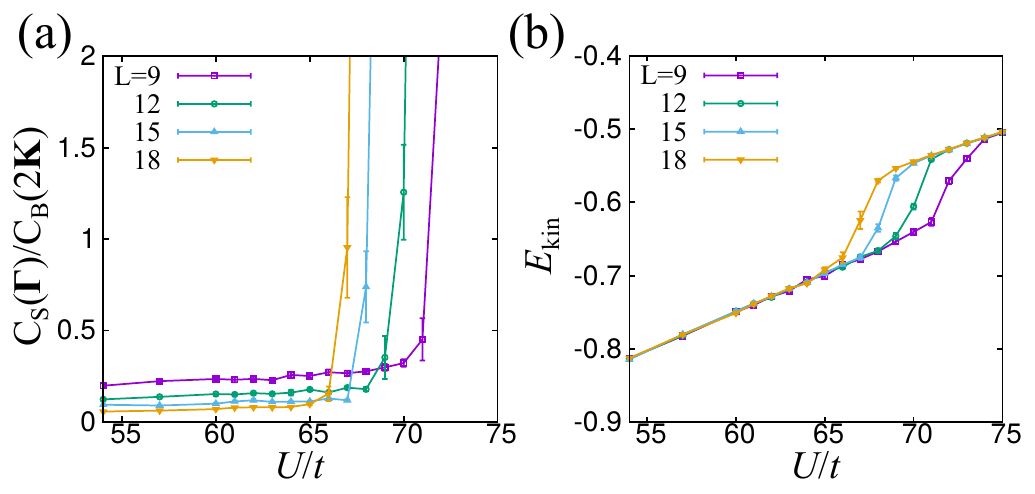}
\caption{ (a) The ratio between structure factor of AB staggered spin correlation at momentum $\boldsymbol{\Gamma}$  and  structure factor of bond correlation at momentum $2\vec{K}$. The sharp jump is one evidence that KVBS to AFMI transition here is first order. (b) Kinetic energy per site of the system. Again the kink in the kinetic energy with tuning parameter $U$ is another evidence of first order transition.
}
\label{fig:kvbs2afm}
\end{figure}

We have also computed the lowest excitation energy for a given momentum. The results are shown in Fig.~\ref{fig:kekuleorder}(e). We see that as $U$ increases, the state at K becomes gapped~\footnote{Rigorously speaking, we mean gap in the thermodynamic limit.}, but the minimum gap remains at K. This suggests that with doping the holes will form pockets at the K points. Since there are two inequivalent K points in the large BZ this predicts that with doping the area of the Fermi pocket should correspond to that of two hole pockets, each with spin degeneracy. However, this contradicts the SdH data on twisted graphene which are consistent with a single doubly degenerate pocket~\cite{cao2018correlated,cao2018unconventional}. If we wish to retain the KVBS state, one option is to see if it is possible to shift the minimum gap to $\Gamma$. We note that a mean field band structure with KVBS order shows a lowest  non-degenerate band at the $\Gamma$ point when $t_3/t<-1$, see Fig.~\ref{fig:kekuleorder}(d). This motivates us to turn on a large and negative $t_3$. We find that the intermediate KVBS phase is quite robust. As shown in the phase diagram,  the third neighbor hopping $t_3$  slowly shrinks its region, but it still occupies quite a large region at $t_3/t=-1.2$. However, we find that the minimum gap remains at the K point (See Fig.~\ref{fig:kekuleorder}(f)).
The trend with further  increase of the magnitude of $t_3$, is to exclude the KVBS phase, ending with a single  phase transition \,---\, from SM phase to AFMI. However, as the magnitude of $t_3$ increases,  the system becomes more like a $\sqrt{3}\times \sqrt{3}$ enlarged lattice model with first neighbor hopping,  making the finite size effect more and more severe. In order to study the phase transition properties of larger $t_3$, especially the way the SM to KVBS and KVBS to AFMI transition meet together, larger system size is needed, which is  beyond computation resources we have currently, and  we leave it for a future study.

The transition from KVBS to AFMI phase here might be a first order transition. The AFMI structure factor is defined as $C_{\text{S}}(\boldsymbol{\Gamma})=\frac{1}{L^{4}}\sum_{ij}\langle(\vec{S}_{\text{A},i}-\vec{S}_{\text{B},i})(\vec{S}_{\text{A},j}-\vec{S}_{\text{B},j})\rangle$, where $\vec{S}_{\text{A/B},i}$ is  spin operator of A/B site in unit cell $i$. As showed in Fig.~\ref{fig:kvbs2afm}(a), the KVBS and AFMI structure factor ratio for different system size does not  cross at a point, but  with singular jump, which is not consistent with a continuous KVBS to AFMI transition, where KVBS and AFMI structure factor ratio may be a renormalization invariant quantity because of emergent SO(5) or SO(4) symmetry at the deconfined quantum critical point~\cite{nahum2015emergent,gazit2018confinement,sato2017dirac}. 
The histogram of KVBS order parameter near KVBS to AFMI transition also does not show any signature of emergent continuous symmetry as showed in Supplemental Material~\cite{suppl}.
In addition, the kinetic energy per site also  shows discontinuous behavior, there is a kink  as showed in Fig.~\ref{fig:kvbs2afm}(b), which is rather like a first order behavior. 

{\it Discussion and Conclusions}\,---\, 
We solved the spinful fermion model with cluster charge interaction on honeycomb lattice by unbiased sign problem free QMC simulation. A $U/t$-$t_3/t$ phase diagram is mapped out. In addition to the well known SM and AFMI phase, we find a KVBS in the intermediate region, which is new and unexpected. We found the transition from SM to KVBS is continuous and belongs to chiral XY universality class. The critical exponents are calculated with high precision, which can be used to benchmark  many analytical methods based on different approximations~\cite{mihaila2017gross,zerf2017four,jian2017fermion,classen2017fluctuation,torres2018fermion,rosenstein1993critical}. 

Interestingly, this KVBS phase is also possible on isotropically strained graphene  by allowing lattice relaxation~\cite{sorella2018structural}, and has already been realized in graphene  grown epitaxially on a copper substrate~\cite{gutierrez2016imaging}.
 Regarding the TBG experiments, the KVBS state is a promising candidate for the correlated insulating phase found in the experiments. Further experiments to search for lattice translational symmetry breaking of the moir\'e pattern will be of great interests. While our results so far do not explain the  single doubly degenerate pocket seen in the experiment, it remains possible that same sublattice hopping may stabilize a single hole pocket at $\Gamma$. Unfortunately such models are extremely hard to be studied by QMC due to the Fermion sign problem.

{\it Acknowledgments}\,---\, 
We thank Liang Fu, T. Senthil and Pablo Jarillo-Herrero for helpful discussions. X.Y.X. and K.T.L. acknowledge the support of HKRGC through Grants No. C6026-16W, No. 16324216, and No.  16307117. 
K.T.L. is further supported by the Croucher Foundation and the Dr Tai-chin Lo Foundation. P.A.L. acknowledges support by DOE under grant FG 02-03ER46076.
The simulation is performed at Tianhe-2 platform at the National Supercomputer Center in Guangzhou.

\bibliography{main}

\clearpage
\onecolumngrid
\appendix
\begin{center}
\textbf{\large Supplemental Material for "Kekul\'e valence bond order in an extended Hubbard model on the honeycomb lattice, with possible applications to twisted bilayer graphene"}
\end{center}
\setcounter{equation}{0}
\setcounter{figure}{0}
\setcounter{table}{0}
\setcounter{page}{1}
\makeatletter

\renewcommand{\thetable}{S\arabic{table}}
\renewcommand{\theequation}{S\arabic{equation}}
\renewcommand{\thefigure}{S\arabic{figure}}

\section{Solving cluster charge model with Determinantal quantum Monte Carlo method}
\label{sec:pqmc}
We use projection version of determinantal quantum Monte Carlo (PQMC)~\cite{blankenbecler1981monte,hirsch1985two,assaad2008world} method to study ground state properties of the cluster charge model
on honeycomb lattice at half-filling. Our model writes as $H=H_t+H_U$ with
\begin{equation}
H_t=-t\sum_{\langle ij \rangle \alpha}c_{i\alpha}^\dagger c_{j\alpha} -t_3 \sum_{\langle\langle\langle ij \rangle\rangle\rangle \alpha}c_{i\alpha}^\dagger c_{j\alpha} + \text{h.c.}
\end{equation}
\begin{equation}
H_U=U\sum_{\hexagon}(Q_{\hexagon}-2)^2
\end{equation}
where the cluster charge is defined as charge per hexagonal plaquette $Q_{\hexagon} \equiv \sum_{i \in \hexagon} \frac{n_i} {3}$ ($n_i=\sum_\alpha n_{i\alpha}$), and we considered both first and third neighbor hopping.

In PQMC, time evolution operator is used to project out the ground state  $|\Psi_{0}\rangle=e^{-\frac{\Theta}{2}H}|\Psi_{T}\rangle$ from a trial wave function $|\Psi_T\rangle$ as $\Theta$ goes to infinity. The physical observables are measured by
\begin{equation}
\langle\hat{O}\rangle=\frac{\langle\Psi_{0}|\hat{O}|\Psi_{0}\rangle}{\langle\Psi_{0}|\Psi_{0}\rangle}=\frac{\langle\Psi_{T}|e^{-\frac{\Theta}{2}H}\hat{O}e^{-\frac{\Theta}{2}H}|\Psi_{T}\rangle}{\langle\Psi_{T}|e^{-\Theta H}|\Psi_{T}\rangle}
\end{equation}
We first perform Trotter decomposition to separate $H_t$ and $H_U$.
\begin{equation}
\langle\Psi_{T}|e^{-\Theta H}|\Psi_{T}\rangle=\langle\Psi_{T}|\left(e^{-\Delta\tau H_{U}}e^{-\Delta\tau H_{t}}\right)^{M}|\Psi_{T}\rangle+\mathcal{O}(\Delta_{\tau}^{2})
\end{equation}
where $\Theta$ is divided into $M$ slices ($\Theta\equiv M\Delta_\tau$). To treat the interaction fermion part, we will use Hubbard Stratonovich (HS) transformation to decouple the interaction part to fermion bilinears coupled to auxiliary fields. We use a fourth order $SU(2)$ symmetric decoupled way
\begin{equation}
e^{-\Delta\tau U(Q_{\hexagon}-2)^{2}}=\frac{1}{4}\sum_{\{s_{\hexagon}\}}\gamma(s_{\hexagon})e^{\alpha\eta(s_{\hexagon})\left(Q_{\hexagon}-2\right)}
\end{equation}
with $\alpha=\sqrt{-\Delta\tau U}$, $\gamma(\pm1)=1+\sqrt{6}/3$,
$\gamma(\pm2)=1-\sqrt{6}/3$, $\eta(\pm1)=\pm\sqrt{2(3-\sqrt{6})}$,
$\eta(\pm2)=\pm\sqrt{2(3+\sqrt{6})}$ and the sum is taken over the auxiliary fields $s_{\hexagon}$ on each hexagon which can take four values $\pm2$ and $\pm1$. After tracing out the free fermions degrees of freedom, we get following formula with a constant factor omitted
\begin{equation}
\langle\Psi_{T}|e^{-\Theta H}|\Psi_{T}\rangle=\sum_{\{s_{\hexagon,\tau}\}}\left[\left(\prod_{\tau}\prod_{\hexagon}\gamma(s_{\hexagon,\tau})e^{-2\alpha\eta(s_{\hexagon,\tau})}\right)\det\left[P^{\dagger}B(\Theta,0)P\right]\right]
\label{eq:mcweight}
\end{equation}
where $P$ is the coefficient matrix of trial wave function $|\Psi_T\rangle$. In the simulation,  we choose the ground state wavefunction of the half-filled non-interacting system (described by $H_t$) as the trial wave function. In the above formula, the $B$ matrix is defined as
\begin{equation}
B(\tau+1,\tau)=e^{V[\{s_{\hexagon,\tau}\}]}e^{-\Delta_\tau K}
\end{equation}
and has properties $B(\tau_3,\tau_1)=B(\tau_3,\tau_2)B(\tau_2,\tau_1)$. Where we have written the coefficient matrix of interaction part as $V[\{s_{\hexagon,\tau}\}]$ and $K$ is the hopping matrix.
The Monte Carlo sampling of auxiliary fields are further performed based on the weight defined in the sum of  Eq.~\eqref{eq:mcweight}. The measurements are performed near $\tau=\Theta/2$. Single particle observables are measured by Green's function directly and many body correlations are measured based on Wick theorem. The equal time Green's function are calculated as
\begin{equation}
G(\tau,\tau)=1-R(\tau)\left(L(\tau)R(\tau)\right)^{-1}L(\tau)
\end{equation}
 with $R(\tau)=B(\tau,0)P$, $L(\tau)=P^{\dagger}B(\Theta,\tau)$. More technique details of PQMC method, please refer to Refs~\cite{blankenbecler1981monte,hirsch1985two,assaad2008world}. 
 For the sign problem free of half-filling case, as we mentioned in the main text, as the model has particle hole symmetry, we can perform a particle hole transformation on down spin ($c_{i\downarrow}^\dagger \rightarrow (-1)^i c_{i\downarrow}$), then the positive $U$ is effectively changed to negative $U$. Then it is easy to check that the Monte Carlo weight defined in  the sum of Eq.~\eqref{eq:mcweight} is always semipositive.
 
\section{Time displaced Green's function and single-particle gap}
To estimate the single-particle gap $\Delta_{\text{sp}}(\vec{k})$, we first measure imaginary-time displaced Green's function $G(\vec{k},\tau)=\frac{1}{2N}\sum_{i,j,a}e^{i\vec{k}\cdot(\vec{R}_i-\vec{R}_j)}\langle c_{i,a}(\frac{\tau}{2})c_{j,a}^\dagger(-\frac{\tau}{2}) \rangle$, where $\langle c_{i,a}(\frac{\tau}{2}) c_{j,a}^{\dagger}(-\frac{\tau}{2}) \rangle$ is given by
\begin{equation}
\langle c_{i,a}(\frac{\tau}{2})c_{j,a}^{\dagger}(-\frac{\tau}{2})\rangle=\frac{\langle\Psi_{T}|e^{-(\frac{\Theta}{2}+\frac{\tau}{2})H}c_{i,a}e^{-\tau H}c_{j,a}^{\dagger}e^{-(\frac{\Theta}{2}-\frac{\tau}{2})H}|\Psi_{T}\rangle}{\langle\Psi_{T}|e^{-\Theta H}|\Psi_{T}\rangle}.
\end{equation}
In above equation, $a$ denotes A/B sublattice, $\vec{R}_i$ is the unit cell coordinate.
During the calculation we should keep $\tau_{\max} \ll \Theta$ and $\Theta \gg 1$ such that both  $e^{-(\frac{\Theta}{2}+\frac{\tau}{2})H}|\Psi_T\rangle$ and $e^{-(\frac{\Theta}{2} - \frac{\tau}{2})H}|\Psi_T\rangle$ are already converged to ground state wave-function. In the calculation of Green's function for system size $L=12$, we find the setting up of $\Theta=26$ and $\tau_{\max}=2$ is enough for a good estimation of single particle gap deep in KVBS phase where the single particle gap is quite large. Finally, the single-particle gap is extracted by the formula $G(\vec{k},\tau)\varpropto e^{-\Delta_{\text{sp}}(\vec{k})\tau} $.
\begin{figure}[h!]
\includegraphics[width=0.6\columnwidth]{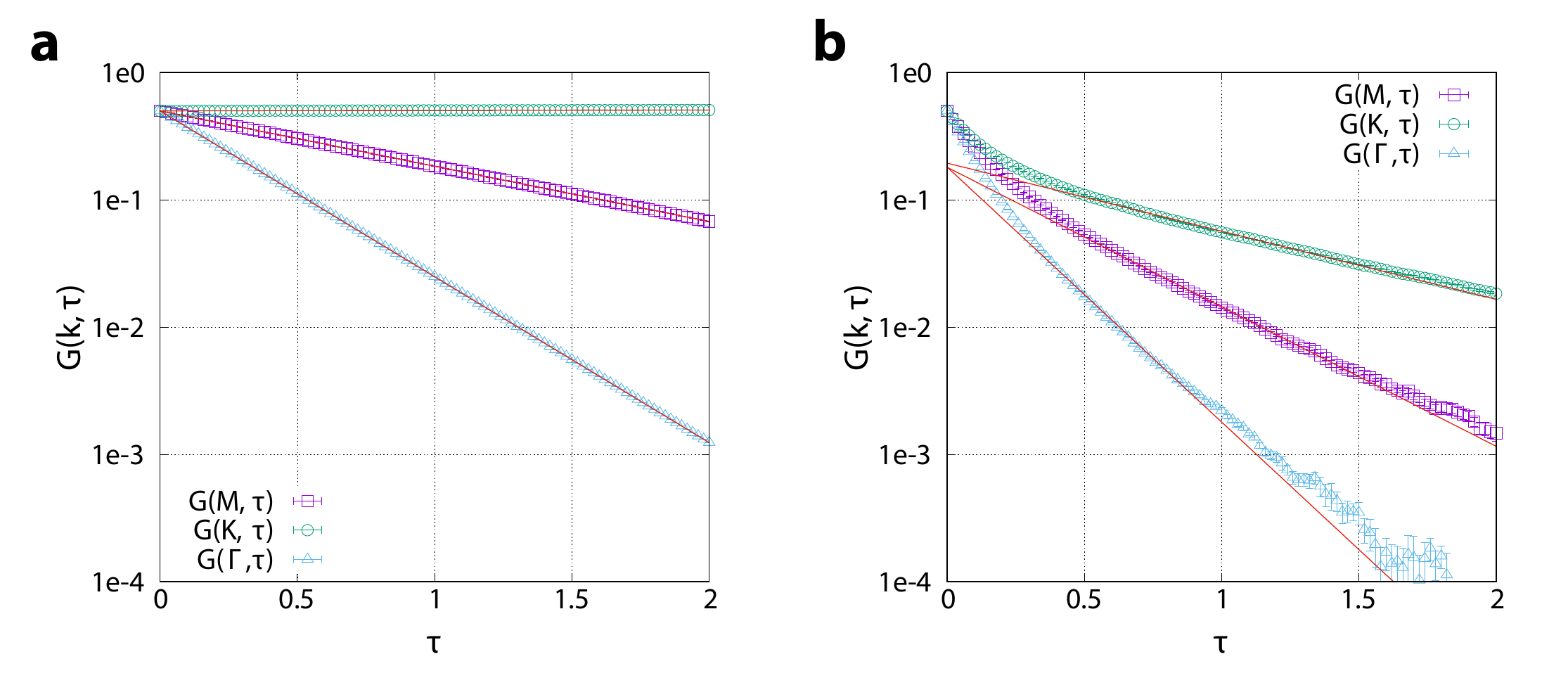}
\caption{Log-plot of imaginary-time displaced Green's function at high symmetry points of $L=12$, $t_3/t=0$ system. \textbf{a}. $U=0$ case. \textbf{b}. $U=24$ case. The absolute value of the slope from linear fitting gives an estimation of the single particle gap $\Delta_{\text{sp}}(\vec{k})$.}
\label{fig:gkt30}
\end{figure}

\section{Numerical instability error at large U}
\begin{figure}[h!]
\includegraphics[width=0.45\columnwidth]{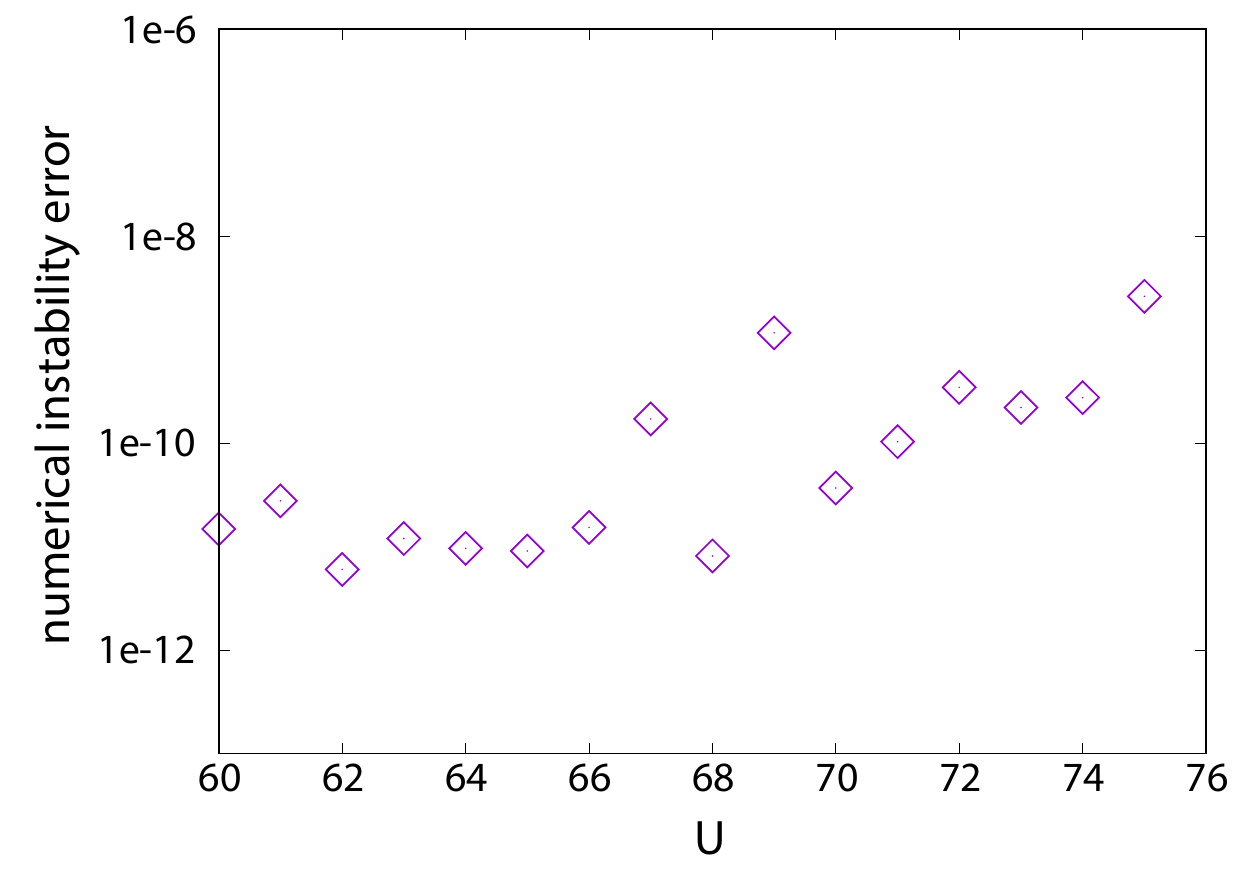}
\caption{Numerical instability error of different $U$ for $L=12$, $t_3/t=0$.}
\label{fig:wraperr}
\end{figure}

As $U$ increases, the condition number of the matrix $B(\Theta,0)$ defined in Sec.~\ref{sec:pqmc} increases accordingly. We use smaller $\Delta_\tau$ at large $U$ cases ($U/t>30$) as we mentioned in the main text. We find to set $\Delta_\tau=0.05$ and perform numerical stabilization every 10 $\Delta_\tau$ steps is enough to obtain error controllable results. We define the numerical instability error as the maximum difference of the matrix elements of equal-time Green's function before and after numerical stabilization. In Fig.~\ref{fig:wraperr} we plot such numerical instability error for different $U$. The numerical instability error increase with $U$, but it is below $10^{-8}$ for all $U$ we studied.

\section{more data for KVBS to AFMI transition}
\begin{figure}[h!]
\includegraphics[width=0.8\columnwidth]{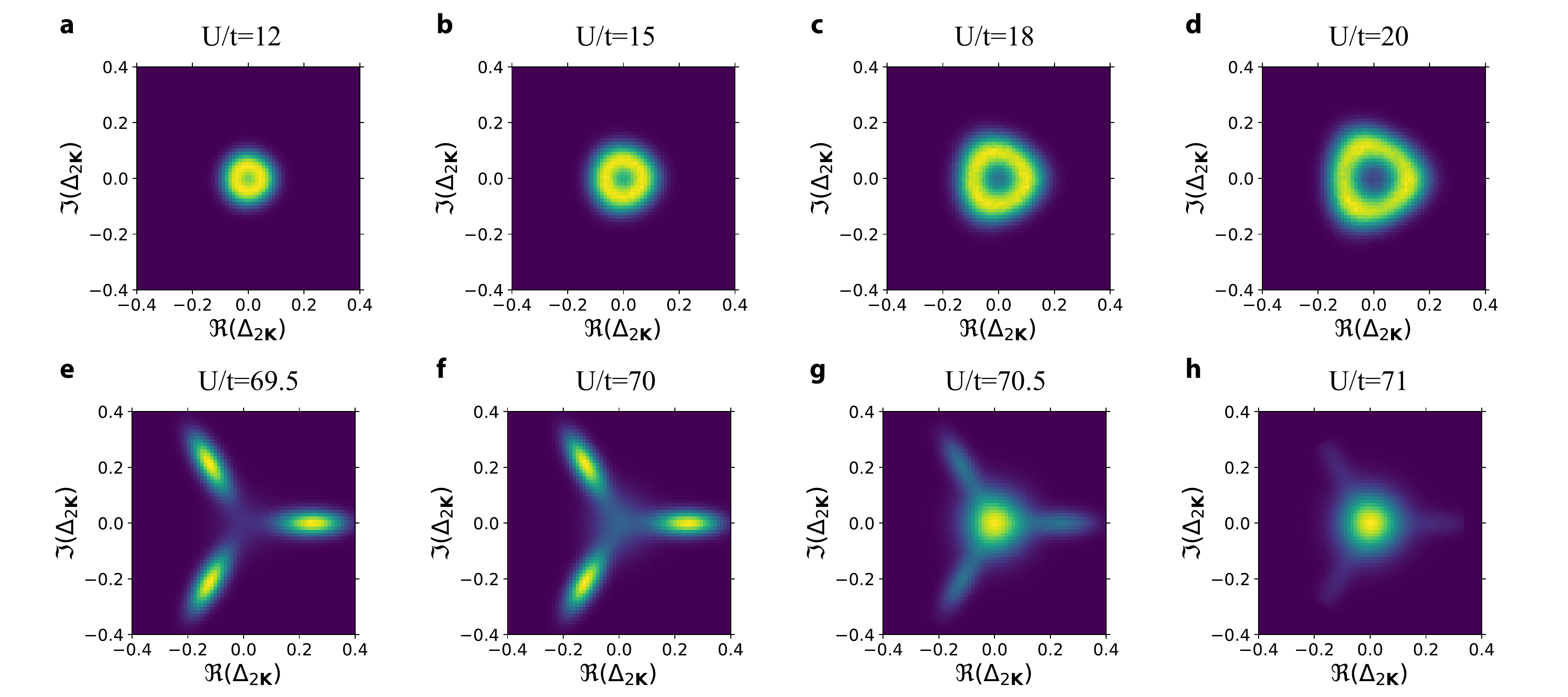}
\caption{\textbf{a}., \textbf{b}., \textbf{c}. \textbf{d}. Histogram of KVBS order parameter near SM to KVBS phase transition. \textbf{e}., \textbf{f}., \textbf{g}. \textbf{h}. Histogram of KVBS order parameter near KVBS to AFMI phase transition. Here $L=12$, $t_3/t=0$. }
\label{fig:histSMtoKVBStoAFMI}
\end{figure}

As we showed in the main text, the KVBS to AFMI transition is very likely to be first order transition. Following are more evidences to support it. First is the histogram of the KVBS order parameter near KVBS to AFMI phase transition. We do not see any signature of emergence of $U(1)$ symmetry as we see at SM to KVBS phase transition. Let's focus on $t_3/t=0$ case. In Fig.~\ref{fig:histSMtoKVBStoAFMI} (e) to (h), we plot the histogram of KVBS order parameter near KVBS to AFMI phase transition point.  In Fig.~\ref{fig:histSMtoKVBStoAFMI} (a) to (d) we also plot the histogram of KVBS order parameter near SM to KVBS phase transition as a comparison.  It is obvious there is no signature of emergent continuous symmetry near KVBS to AFMI phase transition.
\begin{figure}[h!]
\includegraphics[width=0.6\columnwidth]{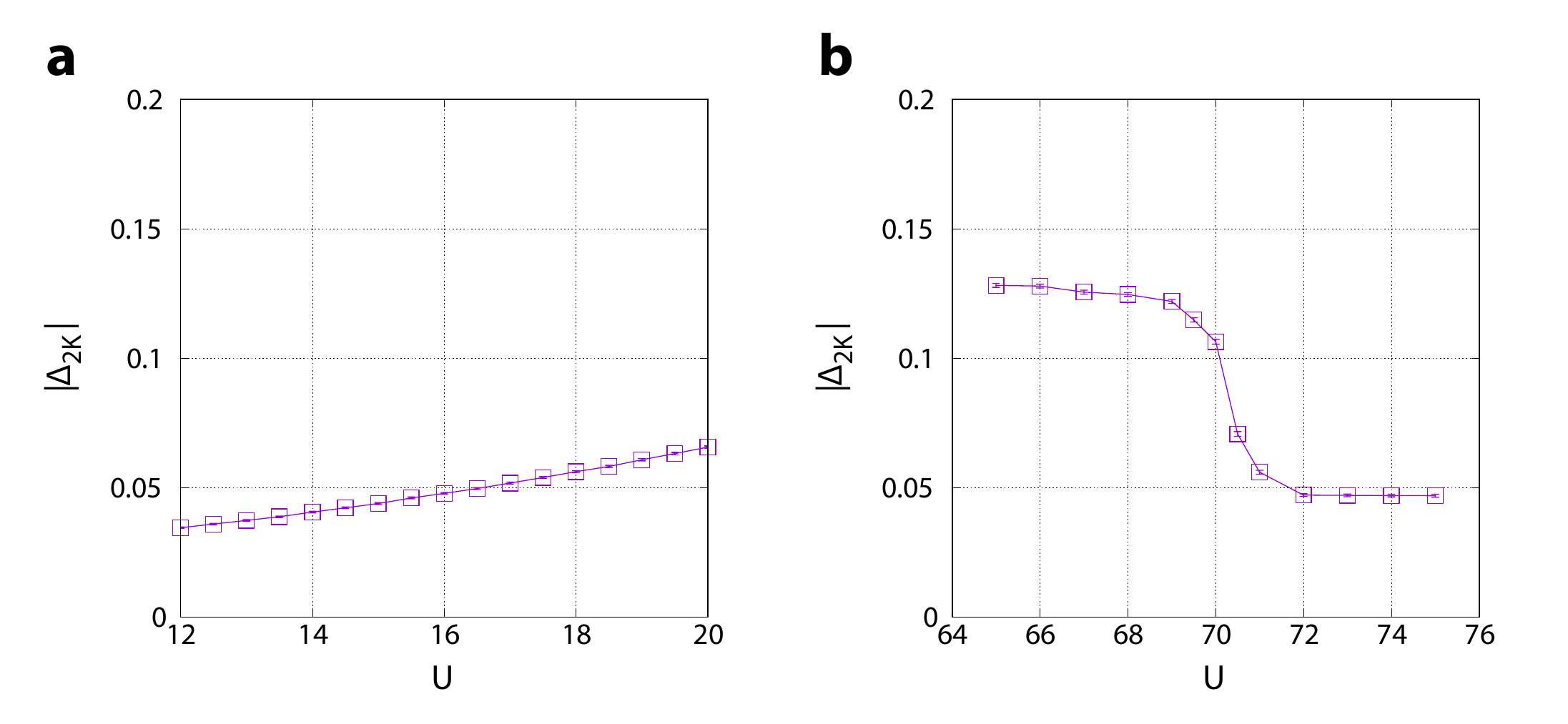}
\caption{\textbf{a}. $|\Delta_{2\mathbf{K}}|$ across SM to KVBS phase transition. \textbf{b}.  $|\Delta_{2\mathbf{K}}|$ across KVBS to AFMI phase transition. Here $L=12$, $t_3/t=0$.}
\label{fig:kvbsordernorm}
\end{figure}
In addition, one can define the strength of the KVBS order parameter as $|\Delta_{2\mathbf{K}}|$ and explore its $U$ dependence across the KVBS to AFMI phase transition as showed in Fig.~\ref{fig:kvbsordernorm}(b), we also see discontinuity, while across the SM to KVBS phase transition, it shows very well continuous behaviour (Fig.~\ref{fig:kvbsordernorm}(a).)

\end{document}